# On the Invariant Integration of a Vector in Some Problems in Mechanics

Saad Bin Mansoor [1, 2]
ORCID: 0000-0002-6227-0386

[1]Department of Mechanical Engineering,
King Fahd University of Petroleum & Minerals, Dhahran, Saudi Arabia

[2] IRC for Renewable Energy and Power Systems (IRC-REPS)
King Fahd University of Petroleum & Minerals, Dhahran, Saudi Arabia

Address: Mail Box 1302, KFUPM. Dhahran. Saudi Arabia.
Telephone: (+966 13) 8608231
Email: saadbin@kfupm.edu.sa

ABSTRACT

Invariant integration of vectors and tensors over manifolds was introduced around fifty years ago by V.N. Folomeshkin, though the concept has not attracted much attention among researchers. Although it is a sophisticated concept, the operation of the invariant integration of vectors is actually required to correctly solve some problems in mechanics. Two such problems are discussed in the present exposition, in the context of a two-dimensional Euclidean space covered by a polar coordinate system. The notion of invariant integration becomes necessary when the space is described without any reference to a Cartesian coordinate system.

Keywords: Invariant integration; mechanics; hydrostatics; centroid

# 1 INTRODUCTION

The operation of "Covariant Integration" was formally introduced by V.N. Folomeshkin in his paper of 1976 [1]. It is concerned with the integration of a vector/tensor field over a manifold as opposed to the integration of a scalar field. The theory of the integration of a scalar field over a manifold is of course well developed. It is quite remarkable that few researchers have appeared to take note of it. In what follows, we will only restrict our discussion to a vector field in two or three-dimensional Euclidean space.

The fundamental problem involved in the integration of a vector field along a curve or over a surface is as follows. While carrying out the operation of integration, we are actually summing an infinite number of vectors that are defined at various points in the space. However, the operation of adding (and multiplying) two or more vectors that are defined at different points in a space is undefined. Only vectors defined at the same point can be summed or multiplied (dot, cross and direct products) together. The solution offered in [1], is to carry out the integration in steps and these are,

1) Select a reference point $P$ anywhere on the curve or surface over which the integration is being carried out and let the covariant basis vectors at that point be $\mathbf{E}_i^P$.
2) Parallel transport the vectors at all points on the region of integration to point $P$ along some path in the Euclidean space, and represent them as a linear combination of the basis vectors $\mathbf{E}_i^P$ (note that in an Euclidean space, the path along which a vector is parallel transported is immaterial).
3) Carry out the integration at point $P$.
4) If required, parallel transport the final result (vector) to some other selected point on the curve or surface and represent it as a linear combination of the basis vectors at that point.

The integration that consists of the previous four steps is termed as 'covariant integration' by Folomeshkin in [1], even though, in the author's estimation, it must actually be termed as 'contravariant integration'. From the previous discussion it becomes clear that the problem of integrating a vector is quite subtle. However, this problem is overlooked possibly because in two common cases of integration involving vectors, the previously described four steps are either not required or are implicitly being carried out. These are,

a) While integrating a scalar composed of a dot product of two vectors in curvilinear coordinates along a curve or a surface. The scalar does not have components and therefore, its integral is well-defined, even though the constituting vectors may change components from point to point.

b) While integrating a vector in Cartesian coordinates. The basis vectors $\hat{\mathbf{i}}$ & $\hat{\mathbf{j}}$ are the same at all points and therefore any vector after parallel transport will have its components unchanged.

It is remarkable that the concept of integrating a vector might also appear in problems of mechanics, and to solve the problem correctly, it is necessary that the concept of 'covariant integration' as introduced in [1] be taken in to account. The problem arises when the region of Euclidean space under study is described by curvilinear coordinates. In the following sections, two problems are discussed, that will serve to illustrate the proper procedure of integrating vectors. Note that vectors are denoted by bold type.

## 2 DETERMINATION OF THE HYDROSTATIC FORCE

Consider the problem of calculating the magnitude and direction of the hydrostatic force on a cylindrical wall of radius $b$ units, as depicted in figure (1). Due to geometrical symmetry of the wall along the $z$ – axis, we only need to consider a single cross-section in the $x-y$ plane. The hydrostatic force on an infinitesimal section of the wall is given as [2],

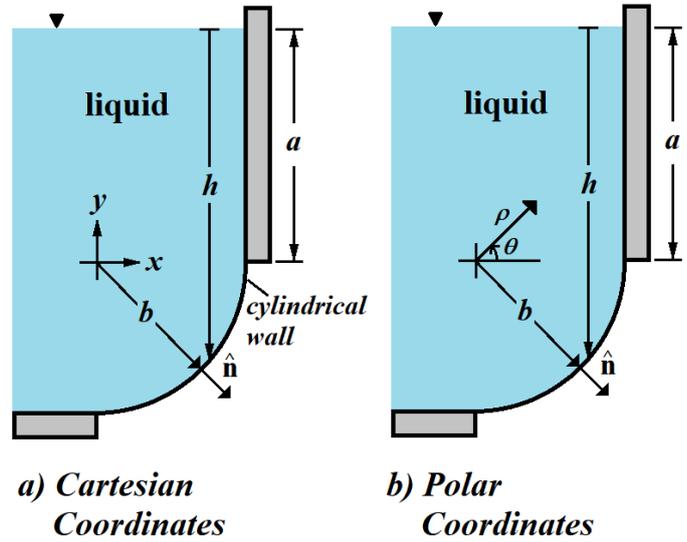

$$d\mathbf{F} = pd\mathbf{A} = \gamma h \hat{\mathbf{n}} dA \qquad (1)$$

In equation (1), $\mathbf{F}$ is the hydrostatic force vector, $p$ is the hydrostatic pressure, $A$ is the total area of the wall, $\gamma$ is the specific weight of the liquid, $h$ is the vertical distance from the free surface to a point on the wall and $\hat{\mathbf{n}}$ is the unit vector perpendicular to the wall at a point. Using the coordinate system shown in figure (1) we can write,

a) Cartesian Coordinates

b) Polar Coordinates

**Fig. 1** Schematic of the hydrostatic problem

$$h = a - b\sin\theta \quad ; \quad \hat{\mathbf{n}} = \cos\theta\,\hat{\mathbf{i}} + \sin\theta\,\hat{\mathbf{j}} \quad ; \quad dA = wbd\theta \qquad (2)$$

where $w$ is the width of the wall in to the paper. The total force is then determined to be,

$$\mathbf{F} = \int_A \gamma h \hat{\mathbf{n}} dA = \int_{3\pi/2}^{2\pi} \gamma(a - b\sin\theta)(\cos\theta\,\hat{\mathbf{i}} + \sin\theta\,\hat{\mathbf{j}})wbd\theta \qquad (3)$$

$$\mathbf{F} = \gamma wb\left[(a + b/2)\hat{\mathbf{i}} - (a + \pi b/4)\hat{\mathbf{j}}\right] \qquad (4)$$

We now would like to solve the problem again, but this time by purely utilizing the polar coordinate system only, without any reference to the Cartesian coordinates or the basis vectors $\hat{\mathbf{i}}\,\&\,\hat{\mathbf{j}}$. To this end, we require the metric tensor and the non-zero connection coefficients, which are considered to be known [3]. These are,

$$[g_{ij}] = \begin{bmatrix} 1 & 0 \\ 0 & \rho^2 \end{bmatrix} \quad ; \quad \Gamma^1_{22} = -\rho \quad ; \quad \Gamma^2_{12} = \Gamma^2_{21} = \frac{1}{\rho} \qquad (5)$$

Then, $h = a - b\sin\theta \quad ; \quad \hat{\mathbf{n}} = \mathbf{E}_1\big/\sqrt{g_{11}} = \mathbf{E}_1 \quad ; \quad dA = wbd\theta \qquad (6)$

$\mathbf{E}_i = \partial \mathbf{r}/\partial x^i$ are the covariant basis vectors in the curvilinear coordinate system that is being utilized to describe the Euclidean space, $\mathbf{r} = \mathbf{r}(\rho,\theta)$ is the position vector and $(x^1, x^2) = (\rho, \theta)$. The hydrostatic force is now to be determined. It is formally given as,

$$\mathbf{F} = \int_A \gamma h \hat{\mathbf{n}} dA = \gamma w b \int_{3\pi/2}^{2\pi} (a - b\sin\theta)\mathbf{E}_1 d\theta \tag{7}$$

The question now is: how to evaluate the integral in equation (7)? $\mathbf{E}_1$ is not a constant but is a function of the angle $\theta$. Moreover, the integral in equation (7) is actually a line integral along the curve $\rho = b$. But more importantly, this integral is in reality undefined. This is because, in the said integral, we are actually summing (an infinite number of) vectors $\mathbf{E}_1$ defined at points all along the curve $\rho = b$. The operation of adding or multiplying two vectors that are defined at two different points is undefined. Only vectors defined at the same point can be added or multiplied. The integral in equation (7) can be given a meaning by implementing the previously described steps.

Let $P$ be a point on the curve $\rho = b, 3\pi/2 \leq \theta \leq 2\pi$ with covariant basis vectors $\mathbf{E}_1^P$ and $\mathbf{E}_2^P$. Since the path of parallel transport is immaterial in an Euclidean space, therefore we can uniquely represent the basis vectors $\mathbf{E}_1$ and $\mathbf{E}_2$ as a linear combination of the vectors $\mathbf{E}_1^P$ and $\mathbf{E}_2^P$. To determine this representation, let us first write,

$$\mathbf{E}_1 = f\mathbf{E}_1^P + \tilde{f}\mathbf{E}_2^P \tag{8}$$

where $f = f(\rho,\theta)$ and $\tilde{f} = \tilde{f}(\rho,\theta)$ are some functions of the variables $\rho \, \& \, \theta$. To proceed,

$$\frac{\partial \mathbf{E}_1}{\partial \rho} = \Gamma_{11}^i \mathbf{E}_i = \cancel{\Gamma_{11}^1} \mathbf{E}_1 + \cancel{\Gamma_{11}^2} \mathbf{E}_2 = \frac{\partial f}{\partial \rho}\mathbf{E}_1^P + \frac{\partial \tilde{f}}{\partial \rho}\mathbf{E}_2^P \tag{9}$$

$$\Rightarrow \frac{\partial f}{\partial \rho} = \frac{\partial \tilde{f}}{\partial \rho} = 0 \Rightarrow f = f(\theta); \tilde{f} = \tilde{f}(\theta) \tag{10}$$

$$\frac{\partial \mathbf{E}_1}{\partial \theta} = \Gamma_{12}^i \mathbf{E}_i = \cancel{\Gamma_{12}^1}\mathbf{E}_1 + \Gamma_{12}^2 \mathbf{E}_2 = \frac{df}{d\theta}\mathbf{E}_1^P + \frac{d\tilde{f}}{d\theta}\mathbf{E}_2^P \tag{11}$$

$$\Rightarrow \mathbf{E}_2 = \rho\left(f_\theta \mathbf{E}_1^P + \tilde{f}_\theta \mathbf{E}_2^P\right) \tag{12}$$

Using the metric tensor definition: $[g_{ij}] = \begin{bmatrix} \mathbf{E}_1 \cdot \mathbf{E}_1 & \mathbf{E}_1 \cdot \mathbf{E}_2 \\ \mathbf{E}_1 \cdot \mathbf{E}_2 & \mathbf{E}_2 \cdot \mathbf{E}_2 \end{bmatrix} = \begin{bmatrix} 1 & 0 \\ 0 & \rho^2 \end{bmatrix}$ we get,

$$1 = f^2 + \rho_P^2 \tilde{f}^2 \; ; \; 0 = ff_\theta + \rho_P^2 \tilde{f}\tilde{f}_\theta \; ; \; 1 = f_\theta^2 + \rho_P^2 \tilde{f}_\theta^2 \tag{13}$$

If we let,

$$f = A\cos(\theta - \theta_P) + B\sin(\theta - \theta_P) \quad ; \quad \rho_P \tilde{f} = A\sin(\theta - \theta_P) - B\cos(\theta - \theta_P) \tag{14}$$

where $A$ & $B$ are numerical constants, then all of equations (13) are identically satisfied, provided that $A^2 + B^2 = 1$. Substituting equations (14) in equations (8) & (12), we get,

$$\mathbf{E}_1 = [A\cos(\theta - \theta_P) + B\sin(\theta - \theta_P)]\mathbf{E}_1^P + \frac{1}{\rho_P}[A\sin(\theta - \theta_P) - B\cos(\theta - \theta_P)]\mathbf{E}_2^P \tag{15}$$

$$\mathbf{E}_2 = \rho_P[-A\sin(\theta - \theta_P) + B\cos(\theta - \theta_P)]\mathbf{E}_1^P + [A\cos(\theta - \theta_P) + B\sin(\theta - \theta_P)]\mathbf{E}_2^P \tag{16}$$

Substituting $\theta = \theta_P$ in equations (15) & (16) we get,

$$\mathbf{E}_1^P = A\mathbf{E}_1^P - \frac{B}{\rho_P}\mathbf{E}_2^P \quad ; \quad \mathbf{E}_2^P = \rho_P B\mathbf{E}_1^P + A\mathbf{E}_2^P \tag{17}$$

$$\Rightarrow \quad A = 1 \quad ; \quad B = 0 \tag{18}$$

So that, finally, we can write,

$$\mathbf{E}_1 = \cos(\theta - \theta_P)\mathbf{E}_1^P + \frac{\sin(\theta - \theta_P)}{\rho_P}\mathbf{E}_2^P \tag{19}$$

$$\mathbf{E}_2 = -\rho_P \sin(\theta - \theta_P)\mathbf{E}_1^p + \cos(\theta - \theta_P)\mathbf{E}_2^p \tag{20}$$

Substituting equation (19) in the integral in equation (7) we get,

$$\mathbf{F} = \gamma w b \int_{3\pi/2}^{2\pi} (a - b\sin\theta)\left(\cos(\theta - \theta_P)\mathbf{E}_1^P + \frac{\sin(\theta - \theta_P)}{\rho_P}\mathbf{E}_2^P\right)d\theta \tag{21}$$

Integral (21) is now well-defined and can be evaluated to get,

$$\mathbf{F} = \gamma w b\left[\{(a + b/2)\cos\theta_P - (a + \pi b/4)\sin\theta_P\}\mathbf{E}_1^P - \frac{1}{\rho_P}\{(a + \pi b/4)\cos\theta_P + (a + b/2)\sin\theta_P\}\mathbf{E}_2^P\right] \tag{22}$$

If we select $(\rho, \theta) = (b, 2\pi)$ to be the point $P$ on the wall, then equation (22) is written as,

$$\mathbf{F} = \gamma w b\left[(a + b/2)\mathbf{E}_1^P - \frac{1}{b}(a + \pi b/4)\mathbf{E}_2^P\right] \tag{23}$$

Moreover, at our selected point $P$, we can also identify the basis vectors to be,

$$\mathbf{E}_1^P = \hat{\mathbf{i}} \quad ; \quad \mathbf{E}_2^P = b\hat{\mathbf{j}} \tag{24}$$

Hence, we can write,

$$\mathbf{F} = \gamma w b \left[ (a+b/2)\hat{\mathbf{i}} - (a+\pi b/4)\hat{\mathbf{j}} \right] \tag{25}$$

The result in equation (25) is the same as in equation (4). In Cartesian coordinates, the basis vectors $\hat{\mathbf{i}} \,\&\, \hat{\mathbf{j}}$ are the same at all points. If we parallel transport all the $\hat{\mathbf{i}} \,\&\, \hat{\mathbf{j}}$ vectors along the curve we will end up with $\hat{\mathbf{i}} = \hat{\mathbf{i}} \,\&\, \hat{\mathbf{j}} = \hat{\mathbf{j}}$. Hence, while evaluating the integral (3) in a Cartesian coordinate system, the steps constituting 'covariant integration' are all being executed implicitly.

## 3   CALCULATION OF THE CENTROID IN POLAR COORDINATES

The centroid $\mathbf{r}_C$ of a region, in two-dimensional Euclidean space, of any shape is given as,

$$\mathbf{r}_C A = \iint_A \mathbf{r} \, d^2 A \tag{26}$$

Let us apply this formula to determine the centroid of a region that is bounded by the spiral $\rho = a\theta$ and by the straight line $\theta = \pi/2$ as shown in the figure (2), but without any reference to the Cartesian coordinates or the basis vectors $\hat{\mathbf{i}} \,\&\, \hat{\mathbf{j}}$. The position vector is given as $\mathbf{r} = \rho \mathbf{E}_1$. The relevant geometric quantities are as described in equation (5). Then,

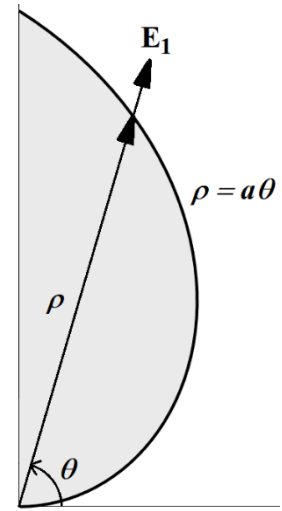

**Fig. 2** Schematic of the centroid problem

$$A = \iint_A d^2 A = \int_0^{\pi/2}\int_0^{\rho} \rho \, d\rho \, d\theta = \frac{1}{2}\int_0^{\pi/2} \rho^2 \, d\theta = \frac{a^2}{2}\int_0^{\pi/2} \theta^2 \, d\theta = \frac{\pi^3 a^2}{48} \tag{27}$$

$$\mathbf{r}_C A = \iint_A \mathbf{r} \, d^2 A = \int_0^{\pi/2}\int_0^{\rho} \rho \mathbf{E}_1 \rho \, d\rho \, d\theta = \frac{1}{3}\int_0^{\pi/2} \rho^3 \mathbf{E}_1 \, d\theta = \frac{a^3}{3}\int_0^{\pi/2} \theta^3 \mathbf{E}_1 \, d\theta \tag{28}$$

In the integral in equation (28), we again face the problem that the integral is undefined as has been discussed earlier. Let $P$ be a reference point with covariant basis vectors $\mathbf{E}_1^P$ and $\mathbf{E}_2^P$ on the spiral. From equations (19) & (20) we write,

$$\mathbf{E}_1 = \cos(\theta - \theta_P)\mathbf{E}_1^P + \frac{\sin(\theta - \theta_P)}{\rho_P}\mathbf{E}_2^P \tag{19}$$

$$\mathbf{E}_2 = -\rho_P \sin(\theta - \theta_P)\mathbf{E}_1^P + \cos(\theta - \theta_P)\mathbf{E}_2^P \tag{20}$$

Substituting equation (19) in the integral in equation (28) we get,

$$\mathbf{r}_C A = \frac{a^3}{2} \int_0^{\pi/2} \theta^3 \left( \cos(\theta - \theta_P) \mathbf{E}_1^P + \frac{\sin(\theta - \theta_P)}{\rho_P} \mathbf{E}_2^P \right) d\theta \qquad (29)$$

Carrying out the integration and using the result in equation (27), we get,

$$\mathbf{r}_C \frac{\pi^3}{16a} = \left[ \left( 6 - 3\pi + \frac{\pi^3}{8} \right) \cos\theta_P + 3\left( \frac{\pi^2}{4} - 2 \right) \sin\theta_P \right] \mathbf{E}_1^P + \frac{1}{\rho_P} \left[ 3\left( \frac{\pi^2}{4} - 2 \right) \cos\theta_P - \left( 6 - 3\pi + \frac{\pi^3}{8} \right) \sin\theta_P \right] \mathbf{E}_2^P \qquad (30)$$

In a Euclidean space, it may not be necessary for the reference point $P$ to lie on the curve or surface along which the integration is being carried out. Therefore, let us select the centroid $C$ to be the point $P$. Then,

$$\mathbf{r}_C \frac{\pi^3}{16a} = \left[ \left( 6 - 3\pi + \frac{\pi^3}{8} \right) \cos\theta_C + 3\left( \frac{\pi^2}{4} - 2 \right) \sin\theta_C \right] \mathbf{E}_1^C + \frac{1}{\rho_C} \left[ 3\left( \frac{\pi^2}{4} - 2 \right) \cos\theta_C - \left( 6 - 3\pi + \frac{\pi^3}{8} \right) \sin\theta_C \right] \mathbf{E}_2^C \qquad (31)$$

However, the centroid is given as, $\mathbf{r}_C = \rho_C \mathbf{E}_1^C$, where $\mathbf{E}_1^C$ is the basis vector at point $(\rho_C, \theta_C)$. Substituting $\mathbf{r}_C = \rho_C \mathbf{E}_1^C$ in equation (31) and simplifying we get,

$$\rho_C = \frac{16a}{\pi^3} \left[ \left( 6 - 3\pi + \frac{\pi^3}{8} \right) \cos\theta_C + 3\left( \frac{\pi^2}{4} - 2 \right) \sin\theta_C \right] \qquad (32)$$

$$0 = \frac{16a}{\pi^3 \rho_C} \left[ 3\left( \frac{\pi^2}{4} - 2 \right) \cos\theta_C - \left( 6 - 3\pi + \frac{\pi^3}{8} \right) \sin\theta_C \right] \qquad (33)$$

From equation (33) we get,

$$\theta_C = \tan^{-1}\left( \frac{6(\pi^2 - 8)}{24(2 - \pi) + \pi^3} \right) \qquad (34)$$

Substituting in equation (32) and simplifying we get,

$$\rho_C = \frac{2a}{\pi^3} \sqrt{\left(24(2-\pi) + \pi^3\right)^2 + 36\left(\pi^2 - 8\right)^2} \qquad (35)$$

From the transformation equations, $x = \rho\cos\theta$ ; $y = \rho\sin\theta$ we get,

$$x_C = \frac{2a}{\pi^3}\left(24(2-\pi) + \pi^3\right) \quad ; \quad y_C = \frac{12a}{\pi^3}\left(\pi^2 - 8\right) \qquad (36)$$

Alternately, let $P$ be the point $(\rho_P, \theta_P) = (a\pi/2, \pi/2)$, then we can write,

$$\mathbf{r}_C = \frac{12a}{\pi^3}\left(\pi^2 - 8\right)\mathbf{E}_1^P - \frac{4}{\pi^4}\left(24(2-\pi) + \pi^3\right)\mathbf{E}_2^P \qquad (37)$$

If we now overlay a Cartesian coordinate system in the usual manner, then we can identify,

$$\mathbf{E}_1^P = \hat{\mathbf{j}} \quad ; \quad \mathbf{E}_2^P = -\frac{a\pi}{2}\hat{\mathbf{i}} \qquad (38)$$

So that, the centroid can also be written as,

$$\mathbf{r}_C = \frac{12a}{\pi^3}\left(\pi^2 - 8\right)\hat{\mathbf{j}} + \frac{2a}{\pi^3}\left(24(2-\pi) + \pi^3\right)\hat{\mathbf{i}} \qquad (39)$$

Or,

$$x_C = \frac{2a}{\pi^3}\left(24(2-\pi) + \pi^3\right) \quad ; \quad y_C = \frac{12a}{\pi^3}\left(\pi^2 - 8\right) \qquad (40)$$

The result in equations (40) can of course be obtained in the customary manner, i.e. by using the Cartesian coordinate system from the beginning.

## 4 CONCLUSIONS

Integrating vector/tensor fields over manifolds could become important in some theoretical or applied problems. The correct procedure of integrating these quantities was expounded around fifty years ago by Folomeshkin and was labelled as 'covariant integration' by him. The importance of this procedure has been illustrated in this paper by applying it to two problems in mechanics. The problems appear to be marginal, however, their purpose is pedagogical. Finally, the author believes that the correct name for what is termed as 'covariant integration' should actually be 'contravariant integration'.

Declarations: Funding and/or competing interests
The author declares no conflicts of interest in this paper

Acknowledgement
The author would like to acknowledge the support of King Fahd University of Petroleum and Minerals (KFUPM), Saudi Arabia. Acknowledgement is also extended to the IRC for Sustainable Energy Systems (IRC-SES), King Fahd University of Petroleum & Minerals, Dhahran, Saudi Arabia